\magnification=1200
\overfullrule=0pt

\baselineskip=12pt

\font\ftit=cmbx10

\parskip=6pt
\parindent=2pc

\font\titulo=cmbx10 scaled\magstep1

\def\section#1{\vskip 1.5truepc plus 0.1truepc minus 0.1truepc
	\goodbreak \leftline{\titulo#1} \nobreak \vskip 0.1truepc
	\indent}

\def\eqdef{\buildrel \rm def \over =}  %"equal by definition"

  %for real numbers
  %for integer numbers
 %for complex numbers \def\covD{{\rmI\!D}} 

%   Si pones en tus definiciones las siguientes instrucciones:

\font\cmss=cmss10
\font\cmsss=cmss10 at 7pt

\def\IZ{\relax\ifmmode\mathchoice
{\hbox{\cmss Z\kern-.4em Z}}{\hbox{\cmss Z\kern-.4em Z}}
{\lower.9pt\hbox{\cmsss Z\kern-.4em Z}}
{\lower1.2pt\hbox{\cmsss Z\kern-.4em Z}}\else{\cmss Z\kern-.4em Z}\fi}

%   podr'as usar  $ \IZ $  para denotar los enteros.

% Hay que usar lo que sigue a continuacion para hacer el caracter
% bold     "  \pmb4   "

\centerline{{\ftit A HAMILTONIAN LATTICE THEORY}} 
\centerline{{\ftit FOR HOMOGENEOUS CURVED SPACETIMES IN 2+1 DIMENSIONS}
\footnote{*}{This work is supported in part by CONACyT grant
400349-5-1714E and by the Association G\'en\'erale pour la
Coop\'eration et le D\'eveloppement \break (Belgium).}}

\vskip 2.5pc

\centerline{A. Criscuolo and H. Waelbroeck${}$}

\vskip 0.5pc

\centerline{Instituto de Ciencias Nucleares, UNAM} \centerline{Apdo.
Postal 70-543, M\'exico, D.F., 04510 M\'exico.} \centerline{e-mail:
hwael@roxanne.nuclecu.unam.mx}

\vskip 1.0pc

\baselineskip=22pt

\vskip 1.5pc

\centerline { Abstract} We propose an exact Hamiltonian lattice 
theory for (2+1)-dimensional spacetimes with homogeneous curvature. 
By gauging away the lattice we find a generalization of the 
``polygon representation'' of (2+1)-dimensional gravity. We compute 
the holonomies of the Lorentz connection 
${\bf A}_i = \omega_i^a {\bf L}_a + e_i^a {\bf K}_a$ and find that the
cycle conditions are satisfied only in the limit $\Lambda \to 0$. 
This implies that, unlike in (2+1)-dimensional Einstein gravity, 
the connection ${\bf A}$ is not flat. 
If one modifies the theory by taking the cycle conditions
as constraints, then one finds that the constraints algebra is
first-class only if the Poisson bracket structure is deformed. 
This suggests that a finite theory of quantum gravity would 
require either a modified action including higher-order curvature
terms, or a deformation of the commutator structure of 
the metric observables.

{\leftskip=1.5pc\rightskip=1.5pc\smallskip\noindent 

	\smallskip}

\vfill\eject

\section{1. Introduction}

 The quest for a lattice version of geometrodynamics which might lead
a discrete proposal for the small scale structure of spacetime has a 
long history. Landmarks include the proposal of 
Ponzano and Regge in $d = 3$~${}^{[1]}$, and 
Penrose's ``spin networks''~${}^{[2]}$. In this note we will 
focus our attention on Hamiltonian lattice theory, where time 
is continuous but {\it space} is discrete. Continuous-time 
Regge calculus was pioneered by Piran and Williams~${}^{[3]}$, 
and Friedmann and Jack~${}^{[4]}$, then developped into the 
``null-strut calculus'' of Miller and Wheeler~${}^{[5]}$ -- 
in these and other efforts, a common difficulty eventually 
limited the potential of Hamiltonian lattice theories both 
for numerical relativity and as fundamental theories for the
small-scale structure problem.

 The difficulty is that the lattice constraints, that are analogous 
to the diffeomorphism constraints of Einstein gravity, fail to 
form a closed algebra under the Poisson brackets: Since these 
constraints are the generators of translations in the continuum 
theory, one says that {\it curvature breaks the translation symmetry} 
in lattice gravity. According to this picture, 
lattice theories describe a patchwork of flat cells, so the curvature 
is concentrated at the lattice ``bones''; the spacetime -- 
and the value of the Einstein action -- will then depend
on the location of the lattice sites, so that one should not
expect translation symmetry to hold. 

 This ``symmetry breaking'' poses serious problems. The counting 
of degrees of freedom differs from the continuum limit, as 
the ``gauge'' degrees of freedom become dynamical at finite lattice
spacing. Also, since the symmetry breaking is very weak, 
the action is nearly constant along the quasi-``gauge orbits'', 
so in computer simulations the integrated history diverges from 
the constraint surface, as integration errors accumulate.

 An example of a lattice theory of geometrodynamics 
with first-class constraints is the exact lattice
formulation of the (2+1)-dimensional gravity for universes with
no continuous matter terms and zero cosmological constant~${}^{[6]}$. 
In this case the Einstein equations imply that spacetime is flat,
so it is not surprising that one can find a lattice theory with 
translation symmetry; the theory can be reduced to only
topological degrees of freedom by fixing the translation 
symmetry~${}^{[7]}$, leading to a representation of spacetime 
as a {\it polygon} in Minkowski space~${}^{[8]}$, which  
provides a convenient framework for canonical quantization~${}^{[9]}$. 

 When the cosmological constant is non-zero, the (2+1)-dimensional
theory with no continuous matter again has only topological degrees 
of freedom, so one hopes that an exact lattice theory can be found.
A first attempt in this direction was based on a complex 
version of the flat lattice theory, where {\it reality conditions}
encode the condition that makes curvature homogeneous~${}^{[10]}$. 
In this way, the constraints are first-class by construction: we 
transposed the difficulty to that of finding reality conditions that 
commute with the constraints. 

 In this Letter, we present an extension of the exact lattice theory 
by introducing a curvature parameter $\Lambda \neq 0$, using only real
variables. This requires introducing additional lattice variables
to specify parallel-transport along links of the lattice; for 
geodesic links in de Sitter space, these are explicit 
functions of the corresponding link vectors. The bracket structure
of the parallel-transport matrices and link vectors together forms 
a deformation of the Poincar\'e algebra.

 The lattice is reduced by gauge fixing, to a ``fundamental polygon''. 
This leaves six constraints which generate global 
$ISO(2,1)$ transformations. We examine the example of the torus universe, 
where the homotopies of the de Sitter connection 
$A = \omega \cdot L + e \cdot K$ are computed explicitly. 

 The solution space of (2+1)-dimensional gravity with a cosmological 
constant is related to the moduli space of flat Lorentz connections,
and this space is spanned by holonomies which satisfy the cycle 
conditions. In the case of the torus, we find that the cycle
conditions are satisfied only to first order in $\Lambda$, so
the connection $A$ is not quite flat. This indicates that the 
solutions are homogeneous spacetimes which satisfy equations 
derived from the Einstein-Hilbert action plus extra terms of higher
order in the curvature.

\vfill\eject

\section{2. The Lattice Theory for $\Lambda = 0$}

 We begin by reviewing the original $\Lambda = 0$ theory~${}^{[6]}$.
The lattice variables are link vectors $E^a(ij)$ ($a = 0, 1, 2$) in 
Minkowski frames at each face ({\it i}), and $SO(2,1)$ matrices 
$M^a_{\ b}(ij)$ for parallel-transport across a link from face ({\it j}) 
to face ({\it i}). The faces are denoted by lower-case lattin 
letters ($i, j, k \cdots$), and the vertices by upper-case letters 
($I, J, K \cdots$)  [Figure 1]. 
These lattice variables have the following Dirac brackets
$$\{ E^a(ij), E^b(ij)\} = \epsilon^{ab}_{\ \ c} \ E^c(ij),
\eqno (2.1)$$
$$\{ E^a(ij), M^b_{\ c}(ij)\} = \epsilon^{ab}_{\ \ d} \ M^d_{\ c}(ij),
\eqno(2.2)$$
$$\{ M^a_{\ b}(ij), M^c_{\ d}(ij) \} = 0. \eqno(2.3)$$
The same variables can be seen from either side of a lattice link;
this leads to the reflection conditions 
$$E^a(ji) = - M^a_{\ b}(ji) E^b(ij),\eqno(2.4)$$
$$M^a_{\ b}(ji) = (M^{-1}(ij))^a_{\ b}.\eqno(2.5)$$
The brackets of variables associated to different links vanish, 
while for the reflected links one finds
$$\{ E^a(ij), E^b(ji)\} = 0,\eqno(2.6)$$
$$\{ E^a(ij), M^b_{\ c}(ji)\} = - M^b_{\ d}(ij)\ \epsilon^{ad}_{\ \ c}.
\eqno(2.7)$$

 The constraints are the face closure relations [Figure 1],
$$J^a(i) \equiv E^a(ij) + E^a(il) + \cdots + E^a(in) \approx 0,
\eqno(2.8)$$
and the flatness conditions for parallel-transport around a lattice
site,
$$P^a(I) \equiv {1 \over 2} \epsilon^{ac}_{\ \ b} \left( {\bf M}(ij)
{\bf M}(jk) \cdots {\bf M}(ni) \right) ^b_{\ c} \approx 0.\eqno(2.9)$$
We denote the number of lattice faces by $N_2$, the number of links 
by $N_1$ and the number of vertices by $N_0$. The $3(N_0 + N_2)$ 
constraints (2.8 - 2.9) satisfy the Poincar\'e-type algebra
$$\{ J^a(i), J^b(i)\} = \epsilon^{ab}_{\ \ c} \ J^c(i),\eqno(2.10)$$
$$\{ J^a(i), P^b(I)\} = \delta(i, I) \ \epsilon^{ab}_{\ \ c} \ 
P^c(I),\eqno(2.11)$$
$$\{ P^a(I), P^b(I)\} = 0,\eqno(2.12)$$
where $\delta(i, I) = 1$ if the vertex $I$ belongs 
to the face $i$ and the constraint $P(I)$ is defined (as above)
by following a counterclockwise loop around $I$ beginning at
the face $i$. If a clockwise loop is followed, $\delta(i, I) = -1$,
and if $I \notin i$, then $\delta(i, I) = 0$.

 The face closure conditions ${\bf J}(i)$ generate $SO(2,1)$ 
transformations of all lattice variables with a Minkowski index 
in face $i$, and the flatness conditions ${\bf P}(I)$ generate 
translations of the lattice vertex:
$$\{ J^a(i), E^b(ij)\} = \epsilon^{ab}_{\ \ c} \ E^c(ij),\eqno(2.13)$$
$$\{ {\bf \xi \cdot P}(I), E^b(ij) \} = - \xi^a.\eqno(2.14)$$

 As defined by the brackets and constraints above, this theory
has $6N_1$ phase space degrees of freedom, minus $3N_0 + 3 N_2$
constraints, which generate the same number of symmetries, so the
number of observable degrees of freedom is
$$6(-N_0 + N_1 - N_2) = -6 \chi = 12g - 12,\eqno(2.15)$$
where $\chi$ is the Euler number of the genus $g$ universe. 
This is also the dimension of the moduli space of flat Lorentz
connections, which results from the Chern-Simons formulation 
of (2+1)-dimensional gravity~${}^{[11, 12]}$.

\vfill\eject

\section{3. Exact Lattice Gravity with $\Lambda \neq 0$}

 Straightforward attempts to introduce curvature in this lattice 
theory lead to the usual difficulty of Hamiltonian
lattice geometrodynamics. One modifies the flatness conditions 
$P^a(I) \approx 0$ by adding to the right-hand side a non-zero term
to reflect the homogeneous curvature, but then finds that the new 
constraints are first class only in the limit of zero curvature, 
$\Lambda \to 0$, or in the continuum limit $\| E \| \to 0$. 

 In this article we choose to avoid this difficulty by giving a 
different interpretation to the parallel-transport matrices ${\bf M}(ij)$.
We will let ${\bf M}(ij)$ denote the parallel transport across
the link ($ij$) {\it at a crossing point arbitrarily near the vertex
I}, where $j$ follows $i$ on a counter-clockwise circle around $I$
(we assume that the universe is orientable). Likewise, 
${\bf M}(ji)$ will denote the reverse crossing but
{\it near the vertex J}. We then introduce matrices to transport 
from a point of face $i$ near the vertex $J$ to a point of face 
$i$ near $I$, ${\bf M}(IJ)$. By convention ${\bf M}(IJ)$ transports
along a segment following the link $ij$ on the side of face $i$,
and vice-versa, ${\bf M}(JI)$ follows the same link in the opposite 
direction ($I \to J$) on the side of face $j$ [Figure 2]. 

 One has the following link-reflection relations.
$$E^a(ij) = - M^b_{\ c}(ij) M^{{-1}^a}_{\ b}(JI) E^c(ji),\eqno(3.1)$$
$${\bf M}(ij) = {\bf M}(IJ) {\bf M^{-1}}(ji) {\bf M}(JI).\eqno(3.2)$$

 With these definitions, the parallel-transport around a vertex $I$
is again trivial, since this is an arbitrarily small loop; thus, the
translation constraints are, as before, 
$$P^a(I) \equiv {1 \over 2} \epsilon^{ac}_{\ \ b} \left( {\bf M}(ij)
{\bf M}(jk) \cdots {\bf M}(ni) \right) ^b_{\ c} \approx 0.\eqno(3.3)$$
Since a lattice face is now a polygon in a curved spacetime, the 
vectors ${\bf E}(ij)$ do not necessarily satisfy face closure conditions. 
On the other hand, we do need constraints that generate $SO(2,1)$ 
conditions and reduce to the face closure conditions as $\Lambda \to 0$. 
In order to write down the correct extension to $\Lambda \neq 0$, we consider
how the structure group $SO(3,1)$ (or $SO(2,2)$) manifests itself 
on the lattice. If one considers geodesic segments in de Sitter
space, the parallel-transport matrices are related to the displacements
by
$${\bf M}(IJ) = exp \left( \sqrt{\Lambda} {\bf E}(ij) \cdot 
{\bf \epsilon} \right) ,\eqno(3.4)$$
where ${\bf \epsilon}^a$ are the $SO(2,1)$ generators, 
$({\bf \epsilon}^a)^b_{\ c} = \epsilon^{ab}_{\ \ c}$.
One finds
$$\{ M^a_{\ b}(IJ), M^c_{\ d}(IJ) \} = {{sin^2 \left( \sqrt{\Lambda} 
\ \| E(ij) \| \right)} \over {\| E(ij) \|^2}} \ \left( \epsilon_{db}
^{\ \ c}\ E^a(ij) - \epsilon^{ac}_{\ \ d}\ E_b(ij) \right) $$
$$+ {{sin \left( \sqrt{\Lambda}\ \|E(ij)\| \right) \left(
1 - cos ( \sqrt{\Lambda}\ \| E(ij) \| ) \right)} 
\over {\| E(ij) \|^3}} \ \left( E_b(ij)\ \delta^a_d E^c(ij)
- \delta^c_b \ E_d(ij) E^a(ij) \right) $$
$$+ {{ \left( 1 - cos ( \sqrt{\Lambda}\ \| E(ij) \| ) \right)^2}
\over {\| E(ij) \|^4}} \ \left( ( \epsilon^a_{\ df}\ 
E^c + \epsilon^{ac}_{\ \ f}\ E_d )  E_b E^f - ( 
\epsilon_{bd}^{\ \ f}\ E^c + \epsilon_b^{\ cf}  E_d )  E_f E^a
\right) .\eqno(3.5)$$
This leads to
$$\{ P^a(IJ), P^b(IJ) \} = \mu\ \Lambda \epsilon^{ab}_{\ \ c} E^c(ij),
\eqno(3.6)$$
where  
$$\mu = {{sin^2 \left( \sqrt{\Lambda} 
\ \| E(ij) \| \right)} \over {\Lambda \| E(ij) \|^2}} .
\eqno(3.7)$$
One has also
$$\{E^a(ij), E^b(ij)\} = \epsilon^{ab}_{\ \ c} E^c(ij),\eqno(3.8)$$
$$\{E^a(ij), P^b(IJ)\} = \epsilon^{ab}_{\ \ c} P^c(IJ),\eqno(3.9)$$
so that ${\bf E}(ij)$ and ${\bf P}(IJ)$ satisfy a deformation of
the Poincar\'e algebra with the deformation parameter $\Lambda$,
where the $SO(2,1)$ generators are ${\bf E}(ij)$. This suggests 
using the same algebraic form of the $SO(2,1)$ constraints as for the 
flat case ($\Lambda = 0$), namely
$$J^a(i) \equiv E^a(ij) + E^a(il) + \cdots + E^a(in) \approx 0.
\eqno(3.10)$$
However, for $\Lambda \neq 0$ these relations do {\it not} imply
that the link vectors represent closed lattice faces in some common 
frame, since one must consider the parallel-transport along links: 
Indeed, the integral of the tangent vector around face $i$ is
$${\bf C}(i) = {\bf E}(ij) + {\bf M}(IJ) {\bf E}(il) 
+ {\bf M}(IJ) {\bf M}(JK) {\bf E}(in) \neq 0, \eqno(3.11)$$
so the faces close only in the limit $\Lambda \to 0$, as ${\bf M}(IJ)
\to {\bf 1}$.

 The curvature for face $i$ is the product of parallel-transport
matrices along the links of the face, e.g.
$${\bf W}(i) = {\bf M}(IJ) {\bf M}(JK) {\bf M}(KI). \eqno(3.12)$$
Expanding to second order in $\sqrt{\Lambda}$, one finds
$$W^a_{\ b}(i) \simeq {\bf 1} + {1 \over 2} \Lambda \epsilon^a_{\ bc}
\left( {\bf E}(il) \wedge {\bf E}(ij) + {\bf E}(in) \wedge {\bf E}(ij)
+ {\bf E}(in) \wedge {\bf E}(il) \right)^c,  \eqno(3.13)$$
where 
$$\left( A \wedge B \right)^c = \epsilon^c_{\ ab} A^b B^a. 
\eqno(3.14)$$
Thus, the curvature is proportional to the area vector of the 
lattice face, to first order in $\Lambda$.

 The link reflection relations simplify if we take into account the
proposed relations between ${\bf M}(IJ)$ and ${\bf E}(ij)$: since
${\bf M}(IJ) {\bf E}(ij) = {\bf E}(ij)$, one finds as before
$$E^a(ij) = - M^a_{\ b}(ij) \ E^b(ji), \eqno(3.15)$$
$${\bf M}(ij) = {\bf M^{-1}}(ji). \eqno(3.16)$$

\

\section{4. Physical Content: Reduction to the Fundamental Polygon}

 As we saw in Section 2, all but a finite number of the degrees of
freedom are pure gauge (for $g \geq 2$ this number is $12 g - 12$ , for the
torus it is equal to 4~${}^{[7]}$). This implies that one can impose 
gauge conditions to reduce the lattice structure to the so-called
``fundamental polygon''. One proceeds as in~${}^{[7]}$: by
translating all lattice sites to a single point, one finds a 
reduced lattice with a single face, one vertex and $2g$ links,
which represent basis elements of the fundamental group. This
reduced lattice can be represented as a polygon with $4g$ edges identified 
in pairs. With the link-reflection relations, one finds that if 
the $2g$ links and matrices are denoted ${\bf E}(\mu)$ and 
${\bf M}(\mu)$ ($\mu = 1, 2, \cdots, 2g$), then the identified 
links are 
$${\bf E}(-\mu) = - {\bf M^{-1}}(\mu) \ {\bf E}(\mu). \eqno(4.1)$$

 Consider the example of the $T^2$ universe [Figure 3]. The
polygon's the edges are ${\bf E}(1)$ and ${\bf E}(2)$ and their 
identified partners are ${\bf E}(-1) = -{\bf M^{-1}}(1) 
{\bf E}(1)$, and ${\bf E}(-2)$. 

 The constraints which generate $SO(2,1)$ transformations are 
$$J^a \equiv \left( {\bf 1 - M^{-1}}(1) \right)^a_{\ b} E^b(1)
+ \left( {\bf 1 - M^{-1}}(2) \right)^a_{\ b} E^b(2) \approx 0,
\eqno(4.2)$$
and the translation constraints require that the parallel-transport 
around the vertex of the polygon is trivial~${}^{[7]}$. 
$$P^a \equiv {1 \over 2}\ \epsilon^{ac}_{\ \ b}\ \biggl(
{\bf M}(1)\ {\bf M^{-1}}(2) \ {\bf M^{-1}} (1) \ {\bf M}(2) 
\biggr) ^b_{\ c} \approx 0.\eqno(4.3)$$
One verifies that these constraints satisfy the algebra
$ISO(2,1)$,
$$\{J^a, J^b\} = \epsilon^{ab}_{\ \ c} J^c,\eqno(4.4)$$
$$\{J^a, P^b\} = \epsilon^{ab}_{\ \ c} P^c,\eqno(4.5)$$
$$\{ P^a, P^b \} = 0.\eqno(4.6)$$

\noindent The Hamiltonian is
$$H = N_a P^a \approx 0, \eqno(4.7)$$
where $N^a$ are the ``lapse-shift'' parameters. One has the 
dynamical equations
$${\buildrel \cdot \over {\bf E}}(1) = ({\bf M}(2) - {\bf 1}) {\bf N}, 
\eqno(4.8)$$
$${\buildrel \cdot \over {\bf E}}(2) = 
{\bf M}(2)({\bf M}^{-1}(1) - {\bf 1}) {\bf N}, \eqno(4.9)$$
$${\buildrel \cdot \over {\bf M}}(1) = 0, \eqno(4.10)$$
$${\buildrel \cdot \over {\bf M}}(2) = 0. \eqno(4.11)$$
Note that this theory is formally equivalent to the $\Lambda = 0$
case; the only change lies in the fact that instead of assuming
that parallel-transport is trivial along an edge of the polygon, 
we are now assuming that it is given by the equivalent of the 
matrix ${\bf M}(IJ) = e^{(\sqrt{\Lambda} {\bf E}(ij) \cdot 
{\bf \epsilon})}$, e.g. by $e^{(\sqrt{\Lambda} {\bf E}(1) \cdot 
{\bf \epsilon})}$, etc.

 In order to examine the relation between this theory and the
Chern-Simons formulation of (2+1)-dimensional gravity, 
we consider the SO(3,1) (SO(2,2)) connection
$${\bf A} = {\bf \omega} \cdot {\bf L} + {\bf e} \cdot {\bf K},$$ 
where $e$ is the dreibein, $\omega$ the spin connection and
$\{ {\bf L}, {\bf K} \}$ are the generators of SO(3,1) (or SO(2,2)):
$$[L^a, L^b] = \epsilon^{ab}_{\ \ c} L^c, \eqno(4.12)$$
$$[L^a, K^b] = \epsilon^{ab}_{\ \ c} K^c, \eqno(4.13)$$
$$[K^a, K^b] = \Lambda \epsilon^{ab}_{\ \ c} L^c. \eqno(4.14)$$

 The variables ${\bf E}(ij)$ and ${\bf M}(IJ)$ are related to the 
integrals of the spin connection and dreibein along the
lattice links as follows ($s \in (0,1)$) 
$$E^a (ij) = \int_I^J N^a_{\ b}(s) \ e^b_{\ i}(s) \ ds^i, \eqno(4.15)$$
$$N^a_{\ b}(s_0) = :e^{\int_I^{s_0} \omega^a_{\ i}(s) 
\ L_a \ ds^i}: \eqno(4.16)$$
$$M^a_{\ b}(IJ) = N^a_{\ b}(s = 1) = :e^{\int_I^J \omega^a_{\ i}(s) 
\ L_a \ ds^i}: \eqno(4.17)$$
From these relations one may show that, with the correct
ordering of the generators, the link-integrals of the connection 
${\bf A}$ give rise to the Lorentz matrices
$${\bf \rho}(IJ) = \ :e^{\int_I^J (\omega^a_{\ i}(s) \ L_a + 
e^a_{\ i}(s) \ K_a) \ ds^i}: $$
$$ \ \ \ \ \ = \ e^{{\bf E} \cdot {\bf K}} \ {\bf M}(IJ) $$
$$ \ \ \ \ \ = \ e^{{\bf E} \cdot ({\bf K + \sqrt{\Lambda} L})} \eqno(4.18)$$
Therefore, the integrals of ${\bf A}$ around two basis loops  $u, v$ 
of the torus are given by
$${\bf \rho}(u) = \ :e^{\oint (\omega^a_{\ i}(s) \ L_a + 
e^a_{\ i}(s) \ K_a) \ ds^i}:$$
$$ \ \ \ \ \ = \ e^{{\bf E}(1) \cdot ({\bf K + \sqrt{\Lambda} L})} 
\ {\bf M}(2), \eqno(4.19)$$
$${\bf \rho}(v) = \ e^{{\bf M}^{-1}(2) {\bf E}(2) \cdot ({\bf K + 
\sqrt{\Lambda} L})} \ {\bf M}^{-1}(1). \eqno(4.20)$$

 Matrices which satisfy the cycle conditions can be identified
with holonomies of a flat connection. Since (2+1)-dimensional
gravity with $\Lambda \neq 0$ is related to a theory of flat 
Lorentz connections, it is interesting to check whether the
matrices ${\bf \rho}(u), {\bf \rho}(v)$ satisfy the cycle
conditions 
$$ {\bf \rho}(u) \ {\bf \rho}(v) \ {\bf \rho}^{-1}(u) \ 
{\bf \rho}^{-1}(v) = {\bf 1} \eqno(4.21)$$
With the constraints, these conditions become
$$e^{{\bf E}(1) \cdot {\cal J}} \ e^{{\bf E}(2) \cdot {\cal J}}
\ e^{-{\bf M}^{-1}(1) \ {\bf E}(1) \cdot {\cal J}} \ 
e^{-{\bf M}^{-1}(2) \ {\bf E}(2) \cdot {\cal J}} = {\bf 1}, \eqno(4.22)$$
where the operators
$${\cal J} = {\bf K} + \sqrt{\Lambda} {\bf L} \eqno(4.23)$$
satisfy the algebra
$$[J^a, J^b] = 2 \sqrt{\Lambda} \epsilon^{ab}_{\ \ c} J^c. \eqno(4.24)$$

 In the limit $\sqrt{\Lambda} \to 0$, one can neglect the commutators 
of the exponential factors in (4.22), so the cycle conditions
become
$$ e^{\left( {\bf E}(1) + {\bf E}(2) - {\bf M}^{-1}(1) \ {\bf E}(1)
- {\bf M}^{-1}(2) \ {\bf E}(2) \right) \cdot {\cal J}} \eqno(4.25)$$
which holds as a consecuence of the constraints 
${\bf J} \approx 0$. 

 This is no longer true if one considers the commutators of the
exponential factors in (4.21):
the $SO(3,1)$ cycle conditions are no longer satisfied, indicating
that the spacetimes described by this theory are not solutions
of Einstein's equations but rather of a different ``gravity theory'',
with corrections proportional to the curvature -- this would be 
equivalent to adding higher-order terms ot the Einstein--Hilbert
action. 

 On the other hand, one may insist that the cycle conditions be
satisfied and the theory be equivalent to a gauge theory of flat
Lorentz connections, and take the cycle conditions (4.22)
as constraints. The constraints ${\bf P} \approx 0$ must hold as well
since the $SO(2,1)$ connection does not have a curvature singularity 
at the vertex, so the cycle conditions turn out to substitute
the constraints ${\bf J} \approx 0$. The new constraints
can be considered to be a deformation ${\bf J'} \approx 0$, where
${\bf J'} = {\bf J} + (\sqrt{\Lambda}-terms)$. 
The requirement that the deformed constraints
${\bf J'} \approx 0$ satisfy the closed (first-class) algebra 
$so(2,1)$ requires deforming the brackets of the polygon variables.
To first order in $\sqrt{\Lambda}$, one finds
$${\bf J'} = {\bf J} + \sqrt{\Lambda} \biggl( 
{\bf E}(1) \wedge {\bf E}(2)
- {\bf E}(1) \wedge {\bf M}^{-1}(1) {\bf E}(1) $$
$$\ \ \ \ \ \ - {\bf E}(1) \wedge {\bf M}^{-1}(2){\bf E}(2) 
- {\bf E}(2) \wedge {\bf M}^{-1}(1){\bf E}(1) $$
$$\ \ \ \ \ \ - {\bf E}(2) \wedge {\bf M}^{-1}(2){\bf E}(2) 
+ {\bf M}^{-1}(1){\bf E}(1) \wedge {\bf M}^{-1}(2){\bf E}(2) 
\biggr) + O(\Lambda). \eqno(4.26)$$
These constraints have the algebra $so(2,1)$ if one assumes the
following deformation of the brackets of any two loop vectors
${\bf V}, {\bf W}$ (e.g., ${\bf V} = {\bf E}(1)$,  
${\bf W} = {\bf M}^{-1}(2) {\bf E}(2)$)
$$\{ V^a, W^b \}_q = \{ V^a, W^b \} + {1 \over 2} (V^a W^b
- V^b W^a). \eqno(4.27)$$
In particular, one has the deformed brackets ($q = 1 + \sqrt{\Lambda}$)
$$\{ E^a(1), E^b(1) \}_q = \epsilon^{ab}_{\ \ c} E^c(1), \eqno(4.28)$$
$$\{ E^a(1), E^b(2) \}_q = {1 \over 2} (q - 1) 
(E^a(1) E^b(2) - E^a(2) E^b(1)), \eqno(4.29)$$
$$\{ E^a(2), E^b(2) \}_q = \epsilon^{ab}_{\ \ c} E^c(2). \eqno(4.30)$$
This implies that the lengths of the two basis loops of the torus
no longer ``commute'': with $l_{\alpha} \eqdef \sqrt{E^2(\alpha)}$, 
one has
$$ \{ l_1, l_2 \} = \| {\bf E}(1) \wedge {\bf E}(2) \|. \eqno(4.31)$$
$$ \ \ \ \ \ \  = l_1 l_2 sin(\theta). \eqno(4.32)$$
Thus, if one insists that the classical phase space should correspond
to the moduli space of flat $SO(3,1)$ connections, one finds that
in the quantum theory the sizes of the two basis loops cannot be
measured simultaneously: Eqn. (4.32) implies that the product 
of the uncertainties on the two lengths will be proportional
to the expectation value of the area of the universe. The algebra
(4.32) is related to the algebra of coordinates on the ``quantum 
plane'', with $\mu - 1 = {ih \over {2\pi}} sin(\theta)$~${}^{[13]}$.

\vfill\eject

\section{5. Discussion}

 This article follows up on a line of research which aims to 
achieve a discrete quantum gravity theory 
by deforming exact lattice versions of topological theories to
introduce local curvature degrees of freedom.  

 In this article, we introduced an exact lattice theory with
{\it homogeneous} curvature. Although this is still a long way from
a lattice theory with inhomogeneous curvature, it shows that 
curvature does not necessarily break the translation symmetry
in a lattice theory of geometry. 

	The variables $E(ij), M(ij)$ satisfy the same equations of motion 
as for $(2+1)$-dimensional gravity with $\Lambda = 0$, so the 
theory is closely related to the Chern-Simons topological invariant 
with $G = ISO(2,1)$. Our result, then, is that the moduli space of 
flat $ISO(2,1)$ connections can be represented as a space of 
homogeneous {\it curved} spacetimes.

 The first-order calculation suggests that it may be possible to 
construct the exact lattice version of the Chern-Simons theory
for the homogeneous gauge groups by taking the cycle conditions
as constraints and deforming the algebra so as to make these 
constraints first-class. The deformed brackets of the loop variables
${\bf E}(\mu)$ then lead  to a non-trivial prediction of the
quantum theory -- that the underlying structure of space is not
a smooth geometry but a deformation thereof, which appears to
be related to the ``quantum space'' structures~${}^{[13]}$.

	The programme, to construct lattice geometrodynamics 
with first class constraints, can proceed from here along 
two different lines: one can either insist
that the lattice theory be a reduction of Einstein's theory to a
finite number of degrees of freedom and deform the algebra of the
finite displacements ${\bf E}(ij)$, or view the Einstein-Hilbert 
action as a low-curvature approximation of the large-scale effective 
action and let the lattice theory provide a prediction of the 
higher-order curvature corrections to the action.

\

\noindent {\bf Acknowledgements} 

\noindent The authors would like to thank Lou Kauffman and
Jos\'e Antonio Zapata for stimulating discussions that led to 
the reactivation of our efforts towards a Hamiltonian lattice
version of geometrodynamics.

\vfill\eject

\centerline{{\ftit References}}

\noindent \item{[1]} Ponzano G and Regge T 1968 {\it Semiclassical
Limit of Racah Coefficients}, in {\it Spectroscopic and Group
theoretical methods in physics}. Amsterdam: North Holland

\noindent \item{[2]} Penrose, R. 1971 {\it Applications of Negative-
dimensional Tensors}, in {\it Combinatorial Mathematics and its 
Applications}, ed. D J A Welsh (Academic Press)

\noindent \item{[3]} Piran T and Williams R M 1986 {\it Phys.
Rev.} D {\bf 33} 1622

\noindent \item{[4]} Friedmann L and Jack I 1986 {\it J. Math. Phys} 
{\bf 27} 2973

\noindent \item{[5]} Kheyferts A, Miller W A and Wheeler J A 
1988 {\it Phys. Rev. Lett.} 2042

\noindent \item{[6]} Waelbroeck, H 1990 {\it Class. Quantum Grav.} 
{\bf 7} 751

\noindent \item{[7]} Waelbroeck, H 1990 {\it Phys. Rev. Lett.}
{\bf 64} 2222

\noindent \item{[8]} Waelbroeck, H 1993 {\it Rev. Mex. Phys.} 
{\bf 39} 831;\ 1991 {\it Nucl. Phys.} B\ {\bf 364} 475

\noindent \item{[9]} Waelbroeck, H and Zertuche, F 1994 {\it
Phys. Rev.} D {\bf 50} 4966;\ {\it Phys. Rev.} D {\bf 50} 4982

\noindent \item{[10]} Waelbroeck, H, Urrutia, L F and Zertuche F
1992 A lattice theory with curvature and exact 
translation symmetry, in {\it Proc. Sixth Marcel Grossman 
Meeting} (Singapore: World Scientific)

\noindent \item{[11]} Witten E 1988 {\it Nucl. Phys.} B {\bf 311} 46

\noindent \item{[12]} Nelson J, Regge T and Zertuche F 1990
{\it Nucl. Phys.} B {\bf 339} 516

\noindent \item{[13]} Takhtajan L.A. 1990: {\it Lectures on Quantum Groups}, 
in: {\it Introduction on Quantum Group and Integrable Massive Models of
Quantum Field Theory}. Edited by M. L. Ge and B. H. Zhao 
(Singapore: World Scientific)

\vfill\eject

\centerline{{\ftit Figure Captions}}

\noindent {\bf Figure 1}. A two-dimensional lattice is locally embedded in
Minkowski space. The embedding is specified by representing the edges 
of each face as link vectors in a frame at that face. For example,
the link vector ${\bf E}(ij)$ (dark arrow) is defined in a frame
at face $i$. Parallel-transport between frames at neighbouring 
faces if given by $SO(2,1)$ matrices ${\bf M}(ij), \cdots$. 
In particular ${\bf E}(ij) = - {\bf M}(ij) {\bf E}(ji)$.

\

\noindent {\bf Figure 2}. For a two-dimensional lattice on a curved 
space, parallel-transport between different vertices of the
same face is non-trivial. Despite this fact, one can still make
use of the variables ${\bf E}(ij), {\bf M}(ij)$, if one assumes
that they are expressed in local frames near the vertex
$I$ ($j$ follows $i$ on a counterclockwise loop around this vertex).
To complete the specification of the connection on the
lattice, we introduce matrices ${\bf M}(IJ)$ which define the
parallel-transport from the local frame at vertex $J$, face $i$
to vertex $I$, face $i$. 

\

\noindent {\bf Figure 3}. The torus can be described as a 
parallelogramme with opposite edges identified. In the ``polygon
representation'', the edges of the parallelogramme are represented
as vectors in a Minkowski frame. However, in the case of 
(2+1)-dimensional gravity with a cosmological constant there is no 
reason to expect that the four vectors in Minkowski space should 
form a closed figure.

\end